\documentclass[sigconf]{acmart}
\AtBeginDocument{%
}

\graphicspath{{figures/}{pictures/}{images/}{./}} 
\usepackage{graphicx} 
\usepackage{algorithm}
\usepackage{algorithmic}
\usepackage{amsmath}   
\usepackage{amssymb}   
\setcopyright{acmlicensed}
\copyrightyear{2025}
\acmYear{2025}
\begin{document}
    \title[Predictability-Aware Motion Prediction for Edge XR]{Predictability-Aware
    Motion Prediction for Edge XR via High-Order Error-State Kalman Filtering } 
\author{Ziyu Zhong}
    \orcid{0009-0005-8513-1230}
    \affiliation{%
      \institution{Lund University}
      \city{Lund}
      \country{Sweden}
    }
    \email{ziyu.zhong@eit.lth.se}
      
\author{Björn Landfeldt}
    \affiliation{%
      \institution{Lund University}
      \city{Lund}
      \country{Sweden}}
\author{Günter Alce}
    \affiliation{%
      \institution{Lund University}
      \city{Lund}
      \country{Sweden}}

\author{Hector A Caltenco}
    \affiliation{%
      \institution{Ericsson}
      \city{Lund}
      \country{Sweden}}


    \begin{abstract}
        As 6G networks are developed and defined, offloading of XR applications
        is emerging as one of the strong new use cases. The reduced 6G latency coupled with edge processing infrastructure will for the first time provide a realistic
        offloading scenario in cellular networks where several computationally intensive
        functions, including rendering, can migrate from the user device and into
        the network. A key advantage of doing so is the lowering of the battery
        needs in the user devices and the possibility to design new devices with
        smaller form factors.

        However, offloading introduces increased delays compared to local execution,
        primarily due to network transmission latency and queuing delays at edge
        servers, especially under multi-user concurrency. Despite the
        computational power of edge platforms, the resulting motion-to-photon (MTP) latency negatively impacts user experience. To mitigate this, motion prediction
        has been proposed to offset delays. Existing approaches build on either
        deep learning or Kalman filtering.
        Deep learning techniques face scalability
        limitations at the resource-constrained edge, as their computational
        expense intensifies with increasing user concurrency, while Kalman filtering suffers
        from poor handling of complex movements and fragility to packet loss inherent in 6G's high-frequency radio interfaces.

        In this work, we introduce a context-aware error-state Kalman filter (ESKF)
        prediction framework, which forecasts the user's head motion trajectory to
        compensate for MTP latency in remote XR. By integrating a motion
        classifier that categorizes head motions based on their predictability, our
        algorithm demonstrates reduced prediction error across different motion classes.
        Our findings demonstrate that the optimized ESKF not only surpasses
        traditional Kalman filters in positional and orientational accuracy but
        also exhibits enhanced robustness and resilience to packet loss.
    \end{abstract}




    \maketitle
    \section{Introduction}
    The convergence of 6G networks and cloud-based extended reality (here termed
    'Remote XR' to distinguish from commercial implementations like \textsc{CloudXR}
    \cite{nvidia_cloudxr}) heralds a new era of immersive experiences, enabling
    high-fidelity rendering and computation offloading to overcome local
    hardware limitations. Local processing consumes considerable energy, leading
    to the need for large batteries in standalone XR headsets. Remote XR, by
    leveraging powerful edge or cloud servers, alleviates these constraints,
    enabling broader accessibility and a more sustainable approach to delivering
    immersive XR experiences. However, the shift to Remote XR introduces a new
    set of challenges, particularly in the realm of latency. The MTP latency, defined
    as the time taken from a user's head movement to the corresponding visual update
    on the display, is a critical factor in maintaining immersion and preventing
    cybersickness \cite{Stauffert_Niebling_Latoschik_2020}. The MTP latency is influenced
    by various factors, including network latency, rendering time, and encoding
    time. As the demand for high-quality VR experiences continues to grow, the
    need for low-latency solutions becomes increasingly important. The challenge
    of MTP latency is particularly pronounced in applications that require rapid
    head movements, such as gaming and interactive simulations. In these scenarios,
    even a small delay can lead to significant degradation in user experience, resulting
    in discomfort and cybersickness.

    Extensive studies of VR have been conducted to eliminate these problems, but most of the solutions are studied for local VR, such as time-warping. For remote VR, it has been shown that motion prediction algorithms can be leveraged to compensate for the delay \cite{hou2020motion} \cite{gul2020kalman}.  Researches have been conducted on 360-degree videos in adaptive streaming, while the
    prediction algorithm was designed for choosing which tiles to include in the field of view \cite{Bao} \cite{Qian} \cite{Fuente}. The accuracy of such a type of task can be lower since the predicted
    position is used for choosing part of a stored video file. For applications
    such as gaming and First Person View (FPV) drone streaming, the prediction problem
    becomes more challenging, where the user experience becomes the key that
    determines whether Remote VR can achieve widespread public adoption. The primary
    challenge stems from two factors. First, in Remote VR, predicted poses are
    used directly by the renderer to generate images for the current viewport, demanding
    higher prediction accuracy than traditional streaming applications. Second, users
    typically exhibit more rapid and dynamic head movements during interactive VR
    experiences, such as gaming, compared to passive activities such as watching
    360-degree videos.

    The existing motion prediction algorithms can be broadly categorized into two
    groups: filter-based and learning-based methods. Filter-based methods, such
    as Kalman filter (KF) \cite{gul2020kalman}, are computationally efficient and can provide accurate
    predictions in real-time applications. However, KF often relies on linear motion
    models \cite{linear_kf}, which may not accurately capture the complex and
    non-linear head movements typical in gaming and interactive simulations. Learning-based
    methods, particularly those relying on deep learning models like LSTMs, have
    shown good accuracy in pose prediction tasks \cite{illahi_learning_2023}. However,
    these methods are computationally intensive, making them less suitable for real-time
    applications in resource-constrained environments at the edge.

    Despite the existence of extensive research, as mentioned in the related work section of \cite{gul2020kalman}, on pose prediction for compensating MTP latency,
    three important research gaps remain underexplored. First, the influence of
    different head motion patterns—especially abrupt, irregular, or highly
    dynamic movements typical in interactive VR applications—on prediction
    accuracy has not been systematically analyzed. Second, the robustness of
    prediction algorithms under real-world network conditions, such as packet loss, is insufficiently addressed, even though these factors can significantly
    degrade system performance. Third, most state-of-the-art deep learning-based
        methods trade accuracy for higher computational resource usage (e.g., GPUs), limiting their scalability and practical deployment on resource-constrained edge platforms. 
    
    To address these challenges, we propose a predictability-aware
    prediction framework that incorporates a motion classifier to demonstrate
    improvements in both prediction accuracy and robustness to packet loss. Experimental results
    show that the high-order ESKF outperforms existing motion prediction algorithms
    in terms of both accuracy and robustness, providing a more effective
    solution for addressing the challenges of MTP latency in EdgeVR
    applications. Crucially, the optimized ESKF operates without the need for
    specialized hardware such as GPUs, making it deployable on cost-sensitive edge
    platforms.

    The remainder of this paper is organized as follows: Section 2 reviews
    related work, Section 3 describes the proposed methodology and predictor design,
    Section 4 presents experimental results and evaluation, Section 5 discusses our
    findings and broader implications, and finally, Section 6 concludes the paper.

    \section{Related Work}
    \subsection{Warping-Based Compensation}
    Asynchronous Timewarping (ATW) is a technique designed to mitigate the effects
    of MTP latency in Virtual Reality (VR) systems. It works by reprojecting the
    last rendered frame based on the most recent head pose data, effectively reducing
    perceived latency. This is achieved by warping the frame to align with
    updated head orientation information, ensuring that the displayed image remains
    consistent with the user’s current viewpoint. \cite{Mark2014} Pose prediction,
    which proactively estimates future head poses (6-DOF position and orientation)
    based on motion sensor data and kinematic models (e.g., Kalman filters \cite{gul2020kalman}
    or deep learning \cite{TRACK} \cite{hou2020motion} \cite{illahi_learning_2023}),
    serves as the foundational pillar for latency reduction in VR systems. By
    generating motion state from predicted poses, it enables early rendering of frames
    that approximate the user's future viewpoint, thereby shifting computational
    burden upstream and significantly compressing the end-to-end Motion-to-Photon
    (MTP) latency pipeline. In contrast, ATW operates reactively: it reprojects existing
    frames using the latest pose data to mitigate latency artifacts after rendering.
    While basic ATW implementations correct only rotational discrepancies (OTW),
    advanced variants like Positional Timewarp (PTW) further address translational
    errors by leveraging depth buffers. \cite{Oculus2016} Crucially, both ATW and
    PTW depends on pose prediction to provide the initial frame for reprojection.
    Their role is complementary—they act as safety nets for residual latency 
    but cannot compensate for errors beyond the scope of the rendered content or
    in dynamic scenes.
    For applications such as collaborative VR that enable geographically separated
    users to interact in a shared virtual space, pose prediction enhances realism
    and reduces perceptual delay, which is critical for maintaining a sense of
    presence and immersion. This is particularly important in applications where
    rapid head movements and interactions are common. Therefore, for a comprehensive
    and reliable MTP latency compensation strategy, pose prediction must operate
    in tandem with ATM to ensure that both rotational and translational errors
    are effectively addressed.

    \subsection{Pose prediction for RemoteXR}
    To address motion extrapolation in latency-constrained RemoteXR environments,
    recent studies advocate LSTM-driven pose prediction frameworks,
    demonstrating efficacy in reducing end-to-end latency while maintaining
    prediction accuracy \cite{hou2020motion} \cite{gul2020kalman}. A key limitation
    of this method is its reliance on GPU-intensive deep learning models, making
    it less efficient for real-time applications compared to lightweight, filter-based
    prediction methods that offer faster, more predictable performance with
    lower computational overhead.  In contrast, the filter-based
    method \cite{gul2020kalman} is computationally lightweight and can operate
    efficiently on CPUs, making them more energy-efficient and practical for real-world
    applications. Therefore, in this work, we focus on improving filter-based
    methods, specifically the Kalman filter (KF), to enhance their performance
    in latency-sensitive remote XR applications.

    \cite{gul2020kalman} proposed a KF-based approach for motion prediction and
    compared the accuracy of prediction against different horizons. This information
    reveals how much latency can be tolerated by users when applications are offloaded
    remotely, which is crucial for researchers to design systems that balance
    computational offloading with user experience, ensuring that the latency introduced
    by remote processing does not degrade the quality of user interaction in VR environments.

    While prior work has advanced motion prediction, significant challenges
    persist in modelling complex motion patterns under network uncertainties. The
    Kalman Filter (KF)-based method \cite{gul2020kalman} relies on linear motion models that fail to capture
    the highly dynamic, non-linear head movements typical in interactive
    applications, resulting in accuracy degradation during rapid motions that are
    hard to predict. Furthermore, \cite{gul2020kalman} models angular velocity
    using first-order quaternion derivatives in state updates. Though
    computationally efficient, integrating these derivatives employs additive operations
    in vector space, violating the multiplicative nature of the SO(3) rotation group \cite{Solà_2017}.
    This fundamental mismatch causes errors in quaternion composition to
    accumulate over time, inducing drift that necessitates frequent ad hoc normalization.
    Such drift compromises prediction accuracy and undermines long-term
    rotational consistency.

    To address these limitations, this work models orientation updates using Lie algebra
    in SO(3)'s tangent space. Unlike quaternion-based integration, this
    framework encodes incremental rotations as minimal perturbations in so(3),
    then maps them to SO(3) via the exponential map \cite{Solà_2017}. This
    ensures all operations respect SO(3)'s manifold constraints, eliminating
    normalization drift. Crucially, intermediate conversion of quaternions to rotation
    matrices enables rigorous SO(3) operations while avoiding singularities and
    drift inherent in direct quaternion differentiation. The rotation matrix acts
    as a faithful embedding of SO(3), guaranteeing unambiguous orientation propagation, particularly
    vital for large rotations where linearized quaternion updates fail geometrically.
    The proposed ESKF framework is designed to be lightweight and
    computationally efficient, making it suitable for deployment on edge servers
    and other resource-constrained environments. By leveraging the ESKF
    framework, we can achieve motion prediction with higher accuracy without the
    need for specialized hardware like GPUs, making it a practical solution for
    real-time applications in VR environments.

    \subsection{Context-aware Predictability}
    Wu et al.~\cite{Wu_Guan_Mao_Cui_Guo_Zhang_2023} point out that LSTM-based approaches
    face difficulties when dealing with motion trajectories that contain abrupt or
    irregular user actions. In such cases, the unpredictability and short duration
    of these movements often exceed the temporal modelling capabilities of LSTM networks,
    resulting in higher prediction errors for complex motion patterns. To design
    a more robust predictor, we adopt entropy as a means to categorize motion
    patterns and systematically assess prediction accuracy for different motion
    patterns. This approach allows our framework to identify and differentiate between
    segments with varying levels of predictability, supporting more effective
    evaluation of prediction methods in VR contexts.

    Recent work by Rossi et al.~\cite{rossi_correlation_2023} has demonstrated a
    strong correlation between the entropy of user trajectories and the predictability
    of their motion in VR environments. Specifically, users exhibiting highly regular
    navigation patterns tend to have lower trajectory entropy, resulting in more
    predictable movements, while those with higher entropy display less predictable
    behavior. By quantifying the entropy of each motion segment using the Lempel-Ziv
    compression-based estimator proposed in~\cite{rossi_correlation_2023}, our classifier
    categorizes motion into distinct predictability classes.

    \section{Methodology}
    
    \subsection{Predictor Design}
    The following Algorithm~ \ref{alg_eskf} shows the general predictor design.
    \begin{algorithm}[H] \caption{Predictability-Aware ESKF Motion Prediction}
        \begin{algorithmic}
            [1] \STATE \textbf{Input:} Pose measurements from OpenXR
            $\mathbf{z}_{k}$ (position $\mathbf{p}_{0}$, orientation
            $\boldsymbol{q}_{0}$), time step $\Delta t$, \STATE \textbf{Output:}
            Updated state $\hat{\mathbf{x}}_{k|k}$, covariance $\mathbf{P}_{k|k}$
            \STATE \STATE \textbf{Step 1 - Initialization} \STATE Set $\mathbf{x}_{0}$,
            $\delta\mathbf{x}_{0}= \mathbf{0}$,
            $\mathbf{P}_{0}= \mathbf{I}$, $\mathbf{Q}= \mathbf{I}$
            \STATE \STATE \textbf{Step 2 - Motion Classification} 
            \FOR {each chunk $i$ in the pose data} 
            
            \STATE \hspace{1em} Compute
            entropy of head motion:
            $H_{k}\leftarrow \text{Entropy}(\mathbf{z}_{k})$ \STATE \hspace{1em}
            Classify motion based on entropy: $C_{k}\leftarrow \text{Classify}(H_{k})$
            \STATE \textbf{Step 3 - Apply low pass filter to each incoming pose} 
            \STATE \textbf{Step 4 - ESKF Prediction and Correction} 
            
            \FOR {each filtered pose data at time step $k$ in chunk $i$ }

            \STATE \textbf{Step 4a - Prediction} \STATE Reset error state: $\delta\hat{\mathbf{x}}
            _{k|k-1}\leftarrow \mathbf{0}$ 
            
                \FOR{ $k$ + $N$ horizon} \STATE
                Predict nominal state: \STATE \hspace{1em} $\mathbf{p}_{k}\leftarrow
                \mathbf{p}_{k-1}+ v_{k-1}\Delta t + \frac{1}{2}\dot{v}
                _{k-1}\Delta t^{2}+ \frac{1}{6}\ddot{v}_{k-1}\Delta t^{3}$ \STATE
                \hspace{1em} $v_{k}\leftarrow v_{k-1}+ \dot{v}
                _{k-1}\Delta t + \frac{1}{2}\ddot{v}_{k-1}\Delta t^{2}$ \STATE
                \hspace{1em} $\dot{v}_{k}\leftarrow \dot{v}_{k-1}+
                \ddot{v}_{k-1}\Delta t$ \STATE \hspace{1em} $\ddot{v}
                _{k}\leftarrow \ddot{v}_{k-1}$ \STATE \hspace{1em} $\boldsymbol
                {q}_{k}\leftarrow \boldsymbol{q}_{k-1}\otimes \exp\left({\omega}_{k-1}\frac{\Delta t}{2}\right)$ 
                \STATE \hspace{1em} $\omega
                _{k}\leftarrow \omega_{k-1}+ \dot{\omega}_{k-1}
                \Delta t + \frac{1}{2}\ddot{\omega}_{k-1}\Delta t^{2}$ 
                \STATE \hspace{1em} $\dot{\omega}_{k}\leftarrow \dot{\omega}_{k-1}+ \ddot{\omega}_{k-1}\Delta t$ 
                
                \STATE \hspace{1em}
                $\ddot{\omega}_{k}\leftarrow \ddot{\omega}_{k-1}$ 
                
                \STATE Assemble predicted state: 
                \STATE \hspace{1em} $\hat{\mathbf{x}}_{k|k-1}\leftarrow [\mathbf{p}_{k}, \mathbf{v}_{k},
                \mathbf{\dot{v}}_{k}, \mathbf{\ddot{v}}_{k}, \boldsymbol{q}_{k}, \boldsymbol
                {\omega}_{k}, \boldsymbol{\dot{\omega}}_{k}, \boldsymbol{\ddot{\omega}}
                _{k}]^{T}$
                \ENDFOR

            \STATE Compute error states transition matrix:
            \STATE \hspace{1em}  $\mathbf{F}_{k}\leftarrow
            \text{computeErrorStatesTransitionMatrix}(\Delta t)$ \STATE Update error
            covariance matrix:
            \STATE \hspace{1em} $\mathbf{P}_{k|k-1}\leftarrow \mathbf{F}_{k}\mathbf{P}_{k-1}\mathbf{F}
            _{k}^{T}+ \mathbf{Q}$
            
            \STATE Generate a random number $r$ uniformly in $[0,1]$
            \IF{$r >$ target\_drop\_rate}   
            \STATE \textbf{Step 4b -  Correction}
                \STATE Compute measurement Jacobian:
                \STATE \hspace{1em} $\mathbf{H}_{k}\leftarrow
                \begin{bmatrix}
                    \mathbf{I}_{3\times3} & \mathbf{0} & \cdots                                                     \\
                    \mathbf{0}            & \cdots     & -\mathbf{J}_{r}^{-1}(\mathbf{R}(\boldsymbol{\theta}_{k})) & \cdots
                \end{bmatrix}$
                \STATE Compute innovation covariance: $\mathbf{S}_{k}\leftarrow
                \mathbf{H}_{k}\mathbf{P}_{k|k-1}\mathbf{H}_{k}^{T}+ \mathbf{R}_{k}$
                \STATE Compute Kalman gain:
                $\mathbf{K}_{k}\leftarrow \mathbf{P}_{k|k-1}\mathbf{H}_{k}^{T}\mathbf{S}
                _{k}^{-1}$
                \STATE Compute innovation: $\mathbf{y}_{k}\leftarrow \mathbf{z}_{k}-
                h(\hat{\mathbf{x}}_{k|k-1})$
                \STATE Update error state:
                $\delta\hat{\mathbf{x}}_{k}\leftarrow \mathbf{K}_{k}\mathbf{y}_{k}$
                \STATE Composite both nominal state and error state to get true
                state: $\hat{\mathbf{x}}_{k|k}\leftarrow \hat{\mathbf{x}}_{k|k-1}+ \delta
                \hat{\mathbf{x}}_{k}$
                \STATE Update covariance:
                $\mathbf{P}_{k|k}\leftarrow (\mathbf{I}- \mathbf{K}_{k}\mathbf{H}_{k}
                ) \mathbf{P}_{k|k-1}$
            \ENDIF 
        \ENDFOR
    \ENDFOR
    \end{algorithmic}
    \label{alg_eskf}
    \end{algorithm}
    
    In our system, the model operates directly on pose data (position and
    orientation) provided by the OpenXR runtime, which fuses inertial and visual
    sensor inputs to generate render-ready pose estimates, which are available at
    the uplink of the streaming pipeline. This approach is chosen to enhance robustness
    in scenarios where IMU data may not be reliably received and to reduce bandwidth
    requirements, as transmitting all IMU data can be costly and prone to loss
    during network transmission. The position and orientation data are provided by
    the OpenXR runtime, which translates IMU data to poses that can be used
    directly by the rendering engine.
    
    Following the standard ESKF formulation \cite{Solà_2017}, we model the true user
    motion state as the sum of a nominal state (obtained from OpenXR pose data) and an error state that captures model uncertainties and sensor noise, as demonstrated in Algorithm~ \ref{alg_eskf}. Our
    system dynamics are derived based on this ESKF framework, where the nominal state
    evolves according to deterministic kinematics, and the error state accounts
    for stochastic disturbances and modeling imperfections. 
    Since this work mainly focuses on comparing the performance of predictors for system dynamics, the process noise covariance is defined as an identity matrix.
    The algorithm integrates
    a motion classifier that categorizes head motions based on their predictability,
    which is used to evaluate the robustness of predictors under different levels
    of predictability. The system state equations include up to third derivatives
    (jerk) for both position and orientation, enabling the state vector to
    capture higher-order motion dynamics. By modelling up to the third derivative
    for both position and orientation and propagating this model for prediction,
    we essentially assume that jerk stays constant across the prediction horizon. 
    However, in our recorded dataset, both positional and angular jerk are highly
    dynamic and unpredictable. Hence, in experiments, we systematically vary the
    order of included derivatives to assess their impact on prediction accuracy.
    This allows us to evaluate how higher-order modelling improves robustness,
    especially during motions with hard predictability, as detailed in the Experimental
    section.

    \section{Experiments}
    \subsection{Experimental Setup}
    All variations of KF-based predictors are implemented in Python and run on an
    Apple M1 chip (8-core CPU, 16 GB RAM). Motion data were sampled at 100 Hz and
    collected from an Oculus Quest 3 HMD using the open-source remote streaming framework
    ALVR, which provides head and controller position and orientation via the
    OpenXR runtime.

    In our experiments, we set the prediction horizon to less than 100 ms,
    consistent with prior work~\cite{illahi_learning_2023}. This choice reflects
    the latency requirements of current open-source and commercial remote VR systems,
    where maintaining motion-to-photon latency below 100 ms is critical for a
    seamless user experience. Li et al.~\cite{Li_Hsu_Lin_Hsu_2020} further
    report that, while round-trip latencies up to 90 ms have limited impact on user
    experience, factors such as bandwidth constraints (as low as 35 Mbps) and
    high packet loss rates (up to 8\%) can significantly degrade performance.
    Therefore, our evaluation focuses on prediction horizons that are representative
    of practical remote VR deployments.

    A butterworth filter with a cutoff frequency of 5 Hz was applied to the data
    to remove high-frequency noise in real-time before sending it to the predictor
    module for prediction. This choice of cutoff frequency is based on
    physiological studies, which indicate that the predominant frequency of head
    rotation typically ranges up to 5 Hz during natural movements. Frequencies
    above this threshold are likely to represent noise rather than intentional
    motion, as supported by prior research \cite{grossman_frequency_1988}. By
    filtering out these higher frequencies, the Butterworth filter ensures that the
    predictor operates on clean and meaningful motion data, enhancing the
    accuracy of the prediction framework.

    After the filtering process, pose data are divided into chunks, each of
    which is passed to a motion classifier that classifies the motion into three
    classes indicating the predictability of the motion chunk. The classifier computes
    the entropy of the motion data and classifies the motion based on the entropy
    value. The actual entropy of a user's trajectory is estimated using the
    Lempel-Ziv compression algorithm, as described in \cite{rossi_correlation_2023}.
    Let $\mathbf{X}= [x_{1}, x_{2}, \dots, x_{T}]$ represent a trajectory of
    positional points in a discretized space. For a sub-sequence $\mathbf{L}_{t}=
    [x_{t}, x_{t+1}, \dots, x_{t+\lambda_t-1}]$ starting at time $t$ and spanning
    $\lambda_{t}$ time slots, the entropy is computed as

    \begin{equation}
        H(\mathbf{X}) = \frac{1}{T}\sum_{t=1}^{T}\log_{2}\left( \frac{T}{\lambda_{t}}
        \right),
    \end{equation}

    where $\lambda_{t}$ is the length of the shortest sub-sequence starting at $t$
    that does not appear earlier in the trajectory. This entropy measure
    quantifies the regularity and predictability of the user's motion. We use this
    entropy equation to classify each chunk of motion into three categories: low
    entropy (high predictability), medium entropy, and high entropy (low predictability).
    Our results confirm a consistent correlation between the entropy of VR trajectories
    and their prediction error. Motions with highly regular navigation styles
    exhibit low entropy, indicating greater predictability, while those with
    high entropy demonstrate less predictable movements. This correlation underscores
    the effectiveness of our entropy-based classification approach in capturing the
    inherent predictability of user motion patterns.

    \subsection{Evaluation Metric}
    The performance of the proposed PsudoESKF method and the baseline methods (KF
    and ESKF) is evaluated using the following metrics:
    \begin{itemize}
        \item \textbf{Position Error}: The position error is computed as the Euclidean
            norm between the predicted and ground-truth position vectors at each
            time step. Formally, for a sequence of $x$ predictions, the position
            error at time step $i$ is given by:
            \[
                e^{(\text{pos})}_{i}= \left\| \mathbf{p}^{(\text{pred})}_{i}- \mathbf{p}
                ^{(\text{true})}_{i}\right\|_{2}
            \]
            where $\mathbf{p}^{(\text{pred})}_{i}$ and $\mathbf{p}^{(\text{true})}
            _{i}$ denote the predicted and actual position vectors at time step $i$,
            respectively. The overall position error can be reported as the mean
            or median of $\{e^{(\text{pos})}_{i}\}_{i=1}^{x}$.

        \item \textbf{Orientation Error}: The orientation error is measured as the
            geodesic (angular) distance between the predicted and ground-truth orientations,
            represented as unit quaternions. This metric operates directly on the
            rotation group, ensuring results are independent of the chosen
            reference frame (bi-invariant) and free from singularities. This is critical
            for head motion prediction in XR, where the orientation of the head can
            be measured relative to different reference frames (e.g., global
            coordinates, camera view, or body-centred frames). The error is
            computed using the angular metric on the 3-sphere ($S^{3}$):
            \begin{align*}
                \text{Orientation Error}= \min \Big( & \left\| \log\left( \boldsymbol{q}_{\text{pred}}\boldsymbol{q}_{\text{true}}^{-1}\right) \right\|,             \\
                                                     & 2\pi - \left\| \log\left( \boldsymbol{q}_{\text{pred}}\boldsymbol{q}_{\text{true}}^{-1}\right) \right\| \Big)
            \end{align*}
            where $\boldsymbol{q}_{\text{pred}}$ and $\boldsymbol{q}_{\text{true}}$
            are the predicted and ground-truth orientation quaternions, and
            $\log(\cdot)$ is the logarithmic map from $S^{3}$ to its tangent
            space. This metric gives the minimal angular displacement required to
            align the predicted orientation with the ground truth, providing
            stable and reference-frame-invariant error measurements ~\cite{Huynh_2009}.

        \item \textbf{Prediction Horizon}: The prediction horizon is the time interval
            over which the prediction is made. It is measured in milliseconds
            and is defined as the time difference between the predicted motion
            and the actual motion. It is computed as:
            \[
                \text{Prediction Horizon}= t_{pred}- t_{true}
            \]
            where $t_{pred}$ is the time of the predicted motion and $t_{true}$
            is the time of the actual motion.
    \end{itemize}
    The position and orientation errors are computed for each time step in the
    prediction horizon, and the average errors are reported for each method. The
    latency is computed as the time difference between the predicted and actual
    motions, and the prediction horizon is defined as the time interval over which
    the prediction is made. The entropy is computed as the average uncertainty
    of the predicted motion over the prediction horizon. The performance of the proposed
    PsudoESKF method is compared with the baseline methods (KF and ESKF) using these
    metrics to evaluate the effectiveness of the proposed method. The results
    are presented in the following sections, including comparisons of position and
    orientation errors, latency, prediction horizon, and entropy for different
    motion patterns.

    \subsection{Filter Comparison}
    To rigorously assess the effectiveness of the proposed PsudoESKF framework, we
    conduct a comparative evaluation against the baseline KF and
    ESKF methods. The analysis focuses on key performance metrics, including position
    and orientation prediction errors. Results are recorded across different
    motion predictability classes, enabling a comparison of each method's accuracy
    and robustness under varying motion dynamics.

    For all filters, the process noise covariance matrix $Q_{k}$ and measurement
    noise covariance matrix $R_{k}$ are set to identity matrices scaled by 1.
    $Q_{k}$ models system process uncertainty, while $R_{k}$ models sensor measurement
    noise. Both are assumed Gaussian, zero-mean, and independent across state
    variables. This standardization ensures a fair comparison of predictor performance,
    isolating the effect of model structure from noise parameter tuning. Since noise
    characteristics vary by device and environment, we fix these values to focus
    solely on model differences.

    - Kalman Filter (KF): KF is implemented the same way as in \cite{gul2020kalman},
    a linear filter that models the system and measurement processes as linear
    and only includes velocity and angular velocity in its state representation.

    - ESKF: The general design of ESKF based predictor is included in section 3.2.
    It is a nonlinear filter that uses the error state to correct the predicted
    state. The true state is represented as a linear combination of the predicted
    state and the error state. In our experiments, the process model for the ESKF
    includes only velocity and angular velocity in the state vector; position
    and orientation are updated based on these quantities.

    - PsudoESKF: The proposed PsudoESKF method extends the ESKF by estimating the
    derivatives of position (e.g., velocity, acceleration) and orientation (e.g.
    angular velocity and acceleration) from the pose data alone, rather than
    relying on direct measurement of these derivatives from the IMU. These estimates,
    referred to as pseudo-measurements, enable the filter to operate effectively
    when only position and orientation are available. The PsudoESKF method uses
    the same process noise covariance matrix $Q_{k}$ and measurement noise
    covariance matrix $R_{k}$ as the ESKF,

    Three variants of the PsudoESKF method (p2o2,p2o3,p3o3) are evaluated, distinguished
    by the order of derivatives incorporated into the state vector for position and
    orientation. The notation "p2o3" and "p3o3" denotes the inclusion of up to
    the second or third derivative for position or orientation, respectively.
    For instance, the p3o3PsudoESKF includes position, velocity, and acceleration
    for position and includes quaternion, angular velocity, and angular acceleration
    for orientation. This systematic ablation study enables assessment of the
    impact of higher-order motion dynamics on prediction accuracy.

    The choice of derivative order in the state vector fundamentally influences
    prediction accuracy because it determines how well the model can represent
    the underlying motion dynamics. Including higher-order derivatives such as acceleration
    and jerk for position, or angular acceleration and angular jerk for
    orientation, enables the filter to account for rapid changes and non-linearities
    in user movement. For example, if only velocity is modeled, the filter
    assumes constant velocity between updates, which fails to capture sudden accelerations
    or decelerations, leading to lag or overshoot in predictions. By incorporating
    acceleration and higher derivatives, the model can anticipate and adapt to
    these changes, resulting in more accurate and responsive predictions. This
    effect is especially pronounced for orientation, where higher-order derivatives
    allow the filter to better track abrupt rotational changes, such as quick
    head turns, which are very common in VR gaming.

    Moreover, the order of derivatives directly determines the integration
    method used for propagating orientation: higher-order models can use more accurate
    integration methods, reducing numerical errors and drift over longer
    prediction horizons. In our experiments, we compare the performance of the
    p2o2 and p2o3 PsudoESKF methods to evaluate the impact of these higher-order
    derivatives on prediction accuracy. The p2o2 PsudoESKF uses a second-order integration
    method denoted as Zed12, while the p2o3 PsudoESKF employs a third-order
    method denoted as Zed23 for orientation propagation.

    The Zed12 and Zed23 methods are numerical integration schemes for evaluating
    rotational quaternions from angular velocities, based on the `zed` mapping,
    a truncated power series designed to preserve quaternion norm and improve
    computational efficiency over the standard exponential map
    \cite{Zupan_Zupan_2014}. The notation is as follows: $\omega_{0}$ is the
    angular velocity at the start of the interval, $\omega_{1}$ is the angular acceleration,
    $\omega_{2}$ is the angular jerk, $h$ is the integration time step, and
    $[\omega_{0}, \omega_{1}]$ denotes the commutator, defined as $[\omega_{0}, \omega
    _{1}] = \omega_{0}\times \omega_{1}$. Zed12 is a second-order method that
    uses a first-degree polynomial approximation of angular velocity, defined as
    \begin{equation}
        \mathrm{Zed12}= \mathrm{zed}\left(\omega_{0}h + \frac{\omega_{1}h^{2}}{2}
        \right).
    \end{equation}
    In contrast, Zed23 is a third-order method employing a second-degree
    approximation, defined as
    \begin{equation}
        \mathrm{Zed23}= \mathrm{zed}\left(\omega_{0}h + \frac{\omega_{1}h^{2}}{2}
        + \frac{\omega_{2}h^{3}}{3}+ \frac{[\omega_{0}, \omega_{1}] h^{3}}{12}\right
        ) .
    \end{equation}
    where the commutator term $[\omega_{0}, \omega_{1}] = \omega_{0}\times \omega
    _{1}$ captures the interaction between angular velocity and angular acceleration
    when integrating rotations~\cite{Zupan_Zupan_2014}.

    The primary distinction is that Zed12 achieves second-order accuracy with a linear
    approximation, while Zed23 attains third-order accuracy by incorporating
    higher-order terms and the commutator. The `zed` mapping methods offer a favorable
    balance between computational efficiency and integration accuracy \cite{Zupan_Zupan_2014},
    making them well-suited for real-time applications on resource-constrained
    edge servers. For these reasons, we adopt this approach in our framework to
    ensure both robust prediction performance and practical deployability.


    \subsection{Results}

    \subsubsection{Phase Lag and Overshoot}
    \begin{figure}[h]
        \centering
        \includegraphics[width=\columnwidth]{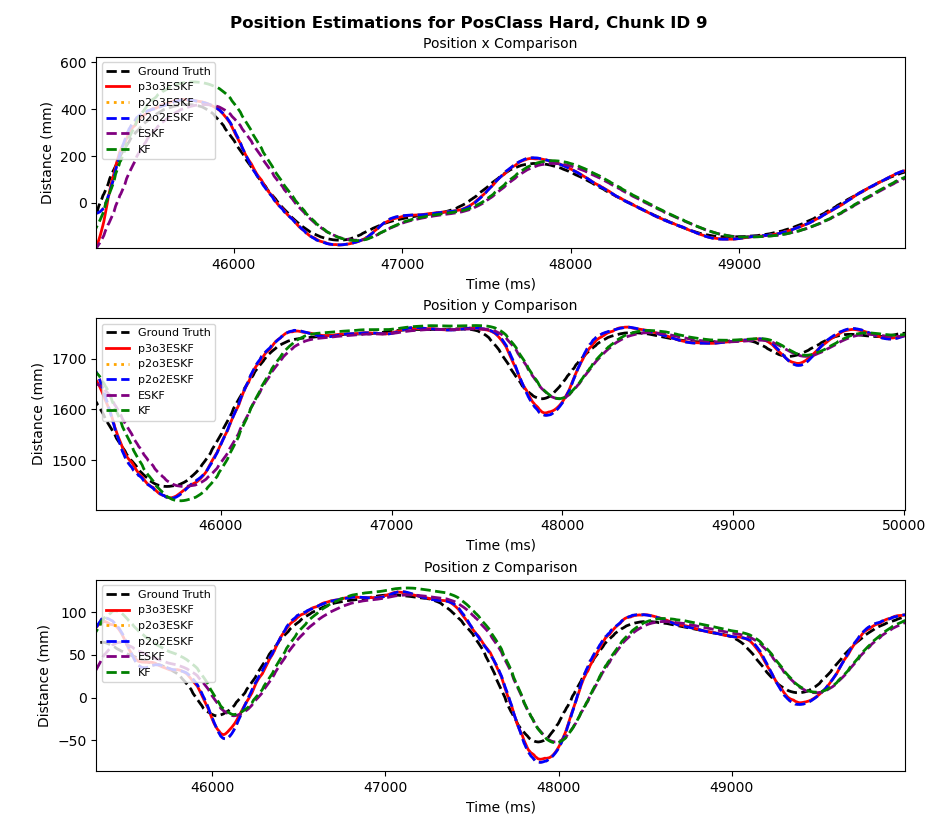}
        \includegraphics[width=\columnwidth]{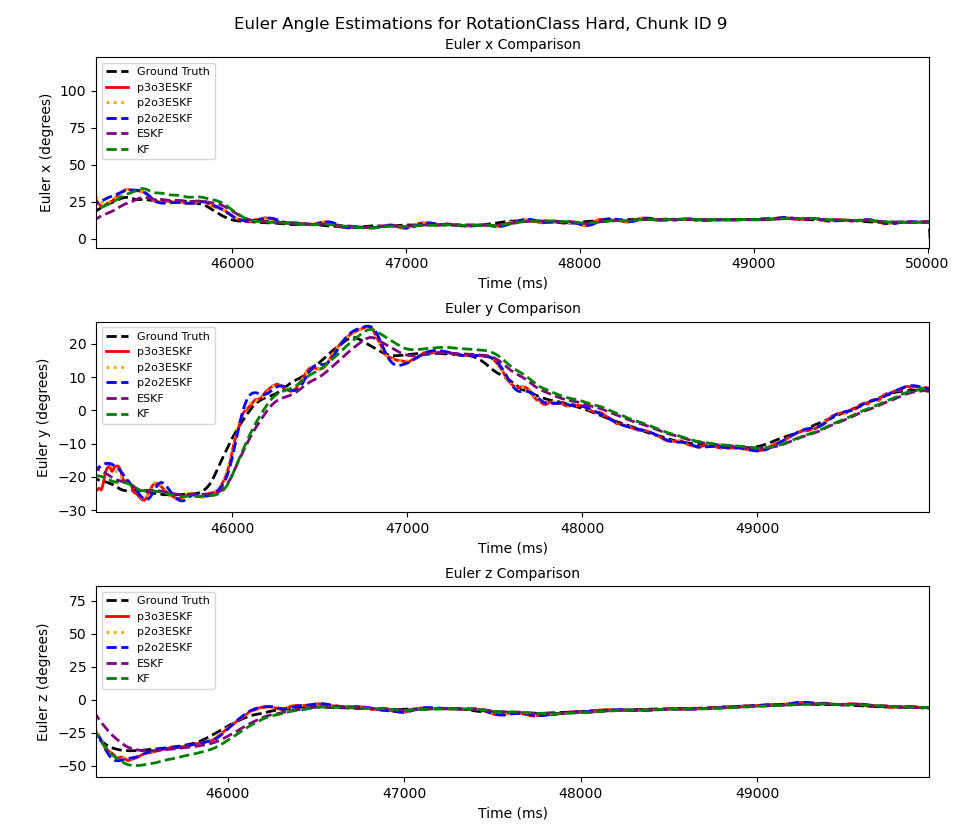}
        \caption{Predicted and ground-truth trajectories of position for x,y,z
        axes in millimeters (top) and orientation for Euler angles in degrees (bottom)
        for a hard motion segment sample.}
        \label{fig:traj_hard}
    \end{figure}

    Figure~\ref{fig:traj_hard} presents a comparative analysis of predicted and ground-truth
    position and orientation trajectories for a representative easy motion segment.
    The results demonstrate that both variants of the proposed PsudoESKF method
    (p2o2 and p2o3) achieve close alignment with the ground truth, exhibiting minimal
    phase lag. In contrast, the baseline methods—Kalman Filter (KF) and ESKF display
    a noticeable phase shift, with predicted trajectories consistently lagging
    behind the ground truth, particularly for the KF. This lag is attributable
    to the KF's reliance on a linear motion model, which is insufficient for capturing
    the non-linear and higher-order dynamics inherent in head motion.

    The ESKF partially mitigates this lag by modelling nonlinearities in the orientation
    update, yet still exhibits a phase shift due to its limited state
    representation. Both PsudoESKF variants further reduce this phase lag by explicitly
    incorporating higher-order derivatives (acceleration and jerk) into the state
    vector, enabling more accurate modelling of rapid changes in user motion. Notably,
    the p2o3 and p2o3 PseudoESKF, which includes up to the third derivative (jerk) for orientation,
    demonstrates superior tracking fidelity, with predicted trajectories closely
    matching the ground truth and exhibiting reduced overshoot compared to the p2o2
    variant.

    Despite the improvements achieved by the proposed PsudoESKF, it is important
    to note a considerable amount of prediction error due to prediction
    overshoot, particularly for hard motion patterns. These errors are most
    pronounced in orientation, where rapid rotational changes challenge even advanced
    predictive models. However, in practice, additional compensation
    techniques—most notably ATW—can be employed to further mitigate the
    perceptual impact of orientation errors. ATW operates by re-projecting the most
    recently rendered frame according to the latest predicted head pose,
    effectively correcting for small to moderate orientation discrepancies that
    arise due to prediction inaccuracies or system latency. This synergy between
    predictive filtering and time warping has been shown to substantially reduce
    motion-to-photon latency and improve visual consistency, especially in
    scenarios with unpredictable head motion.




    \subsubsection{Position and Orientation Errors}
    \begin{figure}[h]
        \centering
        \includegraphics[width=\columnwidth, height=0.2\textheight]{
            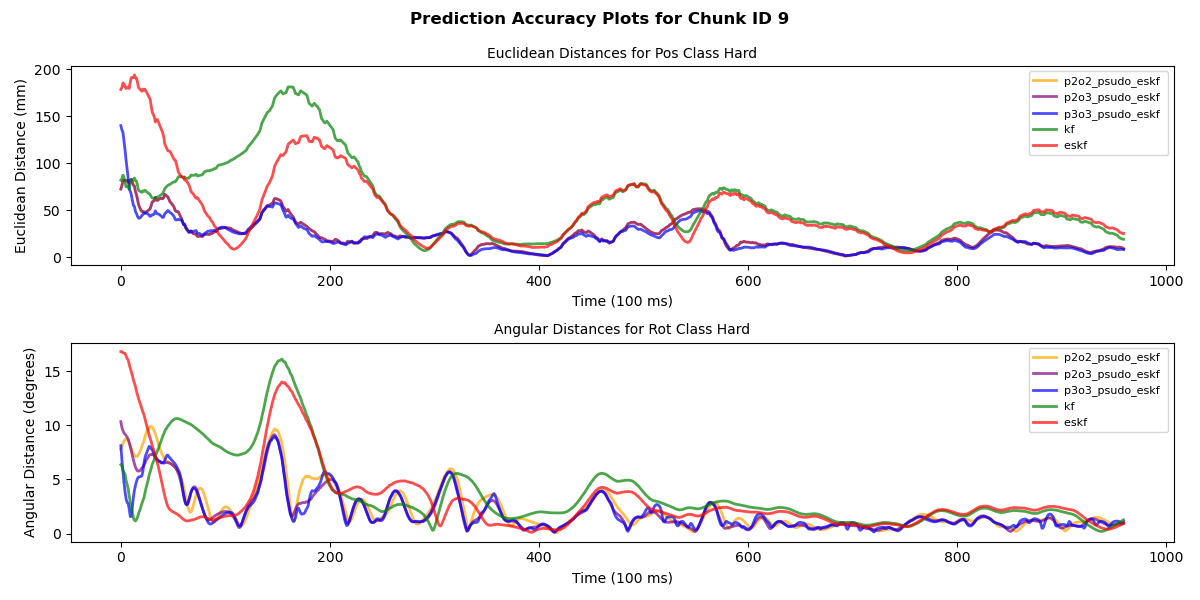
        }
        \caption{Position and Orientation Errors for hard motion patterns}
        \label{fig:mean_error_hard}
    \end{figure}

    Figure~\ref{fig:mean_error_hard} depicts the position and orientation errors
    for hard motion patterns across all methods. The results indicate that the PsudoESKF
    variants (p2o2, p2o3, and p3o3) consistently outperform the baseline methods (KF
    and ESKF) in terms of both position and orientation errors, with the p3o3 variant
    achieving the lowest errors among all. The KF exhibits the
    highest position errors, followed by the ESKF, which, while better than the KF,
    still lags behind the PsudoESKF methods. The position errors for the PsudoESKF
    methods are significantly lower, indicating that the incorporation of higher-order
    derivatives in the state vector leads to more accurate predictions of user
    motion.


    \subsubsection{Different Motion Patterns}

    \begin{table}[h]
        \centering
        \caption{Summary statistics for pose prediction with KF, ESKF, and
        PsudoESKF variants for different head motion patterns. Horizon = 100 ms.}
        \resizebox{\columnwidth}{!}{%
        \begin{tabular}{@{}l l cc cc@{}}
            \toprule \textbf{Model}               & \textbf{Motion Class} & \multicolumn{2}{c}{\textbf{Position (mm)}} & \multicolumn{2}{c}{\textbf{Orientation (deg)}} \\
            \cmidrule(lr){3-4} \cmidrule(lr){5-6} & (Pos, Rot)            & \textbf{Median}                            & \textbf{Mean}                                 & \textbf{Median} & \textbf{Mean}  \\
            \midrule KF                           & (Easy, Easy)          & 2.061                                      & 2.749                                         & 0.973           & 1.177          \\
                                                  & (Medium, Medium)      & 7.120                                      & 7.710                                         & 1.540           & 1.943          \\
                                                  & (Hard, Hard)          & 38.645                                     & 54.096                                        & 2.283           & 3.394          \\
            \midrule ESKF                         & (Easy, Easy)          & 1.943                                      & 2.901                                         & 0.495           & 1.203          \\
                                                  & (Medium, Medium)      & 6.768                                      & 7.303                                         & 1.024           & 1.555          \\
                                                  & (Hard, Hard)          & 35.693                                     & 43.803                                        & 2.057           & 2.725          \\
            \midrule p2o2 PsudoESKF               & (Easy, Easy)          & 1.011                                      & 1.550                                         & 0.441           & 0.831          \\
                                                  & (Medium, Medium)      & 4.162                                      & 5.390                                         & 1.071           & 1.669          \\
                                                  & (Hard, Hard)          & 16.150                                     & 19.286                                        & 1.300           & 1.902          \\
            \midrule p2o3 PsudoESKF               & (Easy, Easy)          & 1.011                                      & 1.550                                         & 0.427           & 0.754          \\
                                                  & (Medium, Medium)      & 4.162                                      & 5.390                                         & 0.937           & 1.415          \\
                                                  & (Hard, Hard)          & 16.150                                     & 19.286                                        & 1.186           & \textbf{1.683} \\
            \midrule p3o3 PsudoESKF               & (Easy, Easy)          & \textbf{0.938}                             & \textbf{1.371}                                & \textbf{0.424}  & \textbf{0.754} \\
                                                  & (Medium, Medium)      & \textbf{3.787}                             & \textbf{4.589}                                & \textbf{0.935}  & \textbf{1.412} \\
                                                  & (Hard, Hard)          & \textbf{15.469}                            & \textbf{17.781}                               & \textbf{1.172}  & 1.711          \\
            \bottomrule
        \end{tabular}%
        } \label{tab:pose_prediction_summary}
    \end{table}

    Table~1 presents an evaluation of the proposed PsudoESKF method in comparison
    with baseline approaches (KF and ESKF) across different motion pattern classes.
    Performance is assessed using both median and mean values of position and
    orientation prediction errors for each motion class. The results demonstrate
    that the PsudoESKF method consistently outperforms the standard Kalman Filter
    (KF) and achieves comparable or superior performance to the ESKF,
    particularly in scenarios involving rapid or unpredictable user movements.

    Notably, the p3o3 variant of PsudoESKF, which incorporates higher-order
    derivatives in the state vector, yields the lowest position and orientation errors
    across all motion classes. This finding underscores the importance of
    modeling higher-order motion dynamics for accurate prediction, especially under
    challenging motion conditions. The systematic reduction in prediction errors
    observed when increasing the order of derivatives from p2o2 to p3o3
    highlights the enhanced capability of the filter to capture complex, non-linear
    user motion.

    Importantly, the PsudoESKF framework is designed to operate using only pose
    data, obviating the need for direct access to raw sensor measurements such as
    IMU data. This approach reduces bandwidth requirements for data transmission
    and simplifies system integration, making it particularly advantageous for edge-based
    VR systems where sensor access may be constrained or subject to network-induced
    delays. By estimating the necessary motion derivatives locally, the PsudoESKF
    method maintains temporal consistency and prediction accuracy even in the presence
    of packet loss or network jitter, thereby providing a robust and efficient
    solution for real-time motion prediction in latency-sensitive VR
    environments.

    \subsubsection{Different prediction horizon}

    \begin{figure}[h]
        \centering
        \includegraphics[width=\columnwidth]{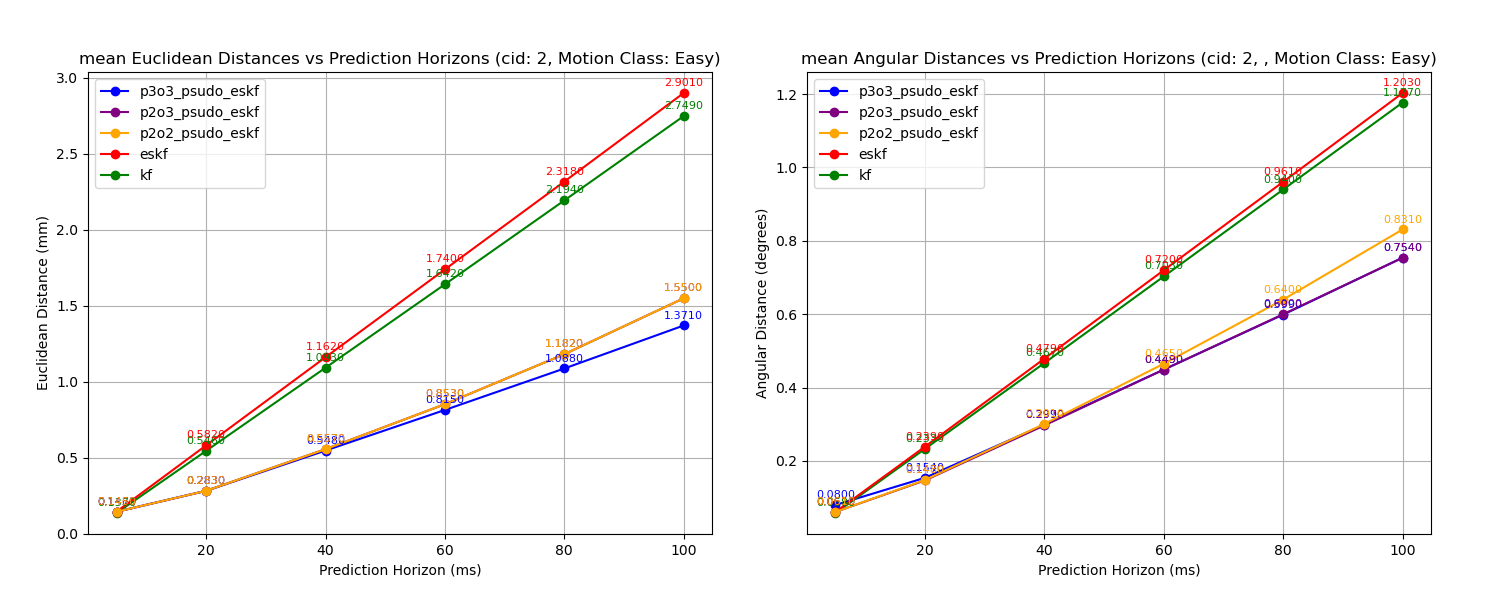}
        \includegraphics[width=\columnwidth]{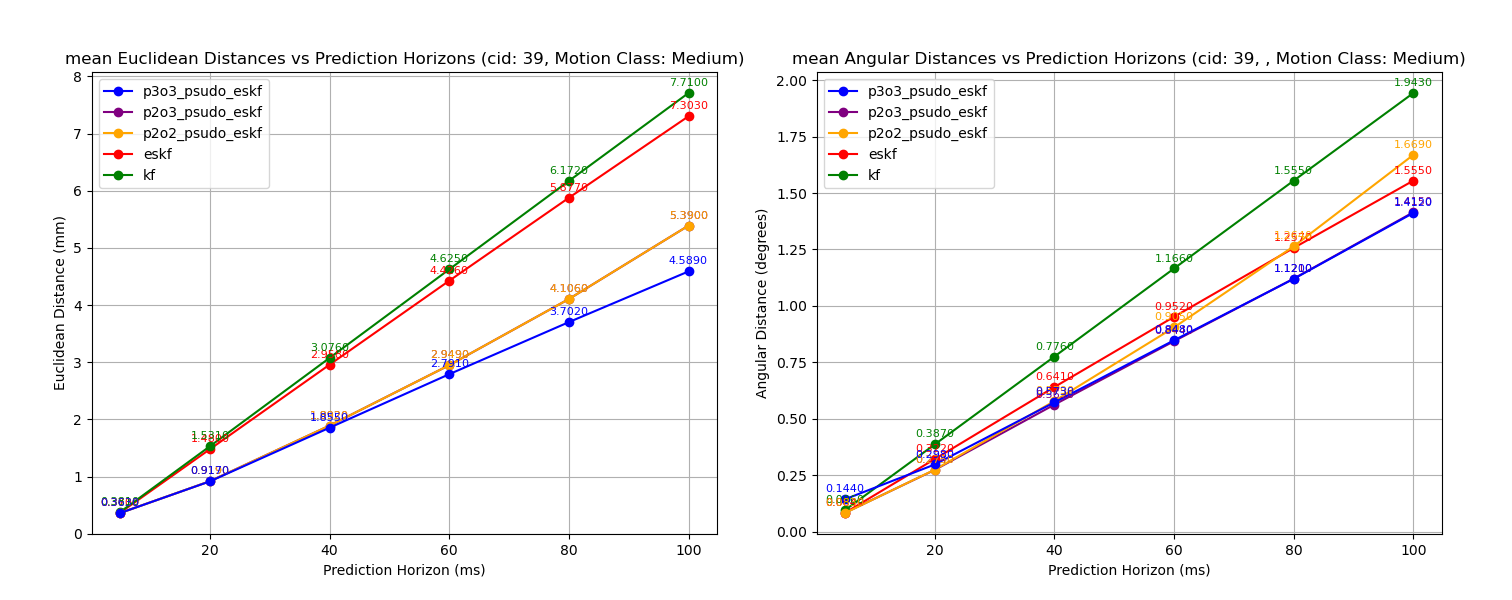}
        \includegraphics[width=\columnwidth]{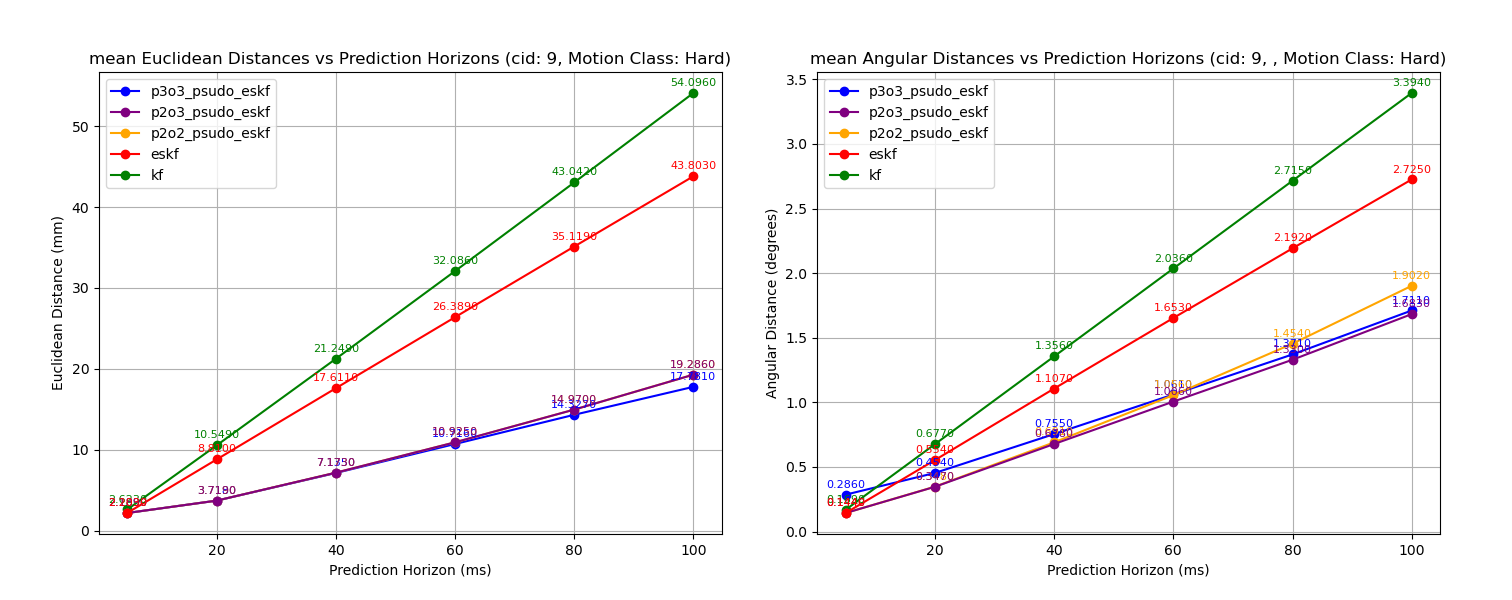}
        \caption{Prediction error (mean) across different prediction horizons
        for (top) easy, (middle) medium, and (bottom) hard motion classes.}
        \label{fig:prediction_error_horizon}
    \end{figure}

    Figure~\ref{fig:prediction_error_horizon} presents a detailed comparison of
    the prediction performance of the proposed PsudoESKF method against the
    baseline KF and ESKF across varying prediction horizons and motion pattern
    classes. The evaluation metrics include both the mean and median of position
    and orientation prediction errors, computed for each prediction horizon.

    The results demonstrate that the PsudoESKF method consistently achieves
    lower prediction errors than both baseline methods, with the performance gap
    widening as the prediction horizon increases. This trend is particularly
    pronounced in the easy motion class, where PsudoESKF maintains minimal error
    growth even at longer horizons, indicating superior temporal stability and
    predictive accuracy. In contrast, both KF and ESKF exhibit a more rapid
    increase in error, reflecting their limited ability to capture higher-order motion
    dynamics and adapt to longer-term predictions.

    For medium and hard motion classes, which are characterized by more abrupt
    and less predictable user movements, the PsudoESKF method continues to outperform
    the baselines. While all methods experience increased errors with longer
    horizons due to the inherent unpredictability of the motion, PsudoESKF demonstrates
    a slower rate of error escalation.

    Overall, these findings highlight the effectiveness of the PsudoESKF
    framework for maintaining prediction accuracy across a range of temporal
    horizons and motion complexities.


    \subsubsection{Different data droprate}

    \begin{figure}[h]
        \centering
        \includegraphics[width=\columnwidth]{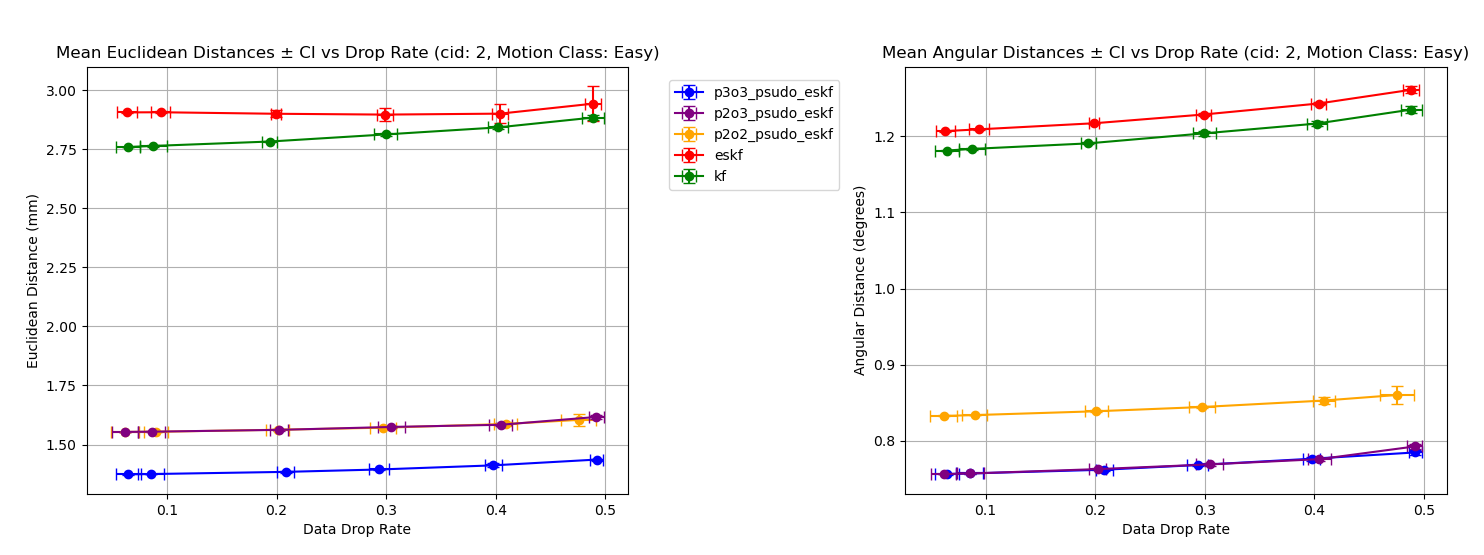}
        \includegraphics[width=\columnwidth]{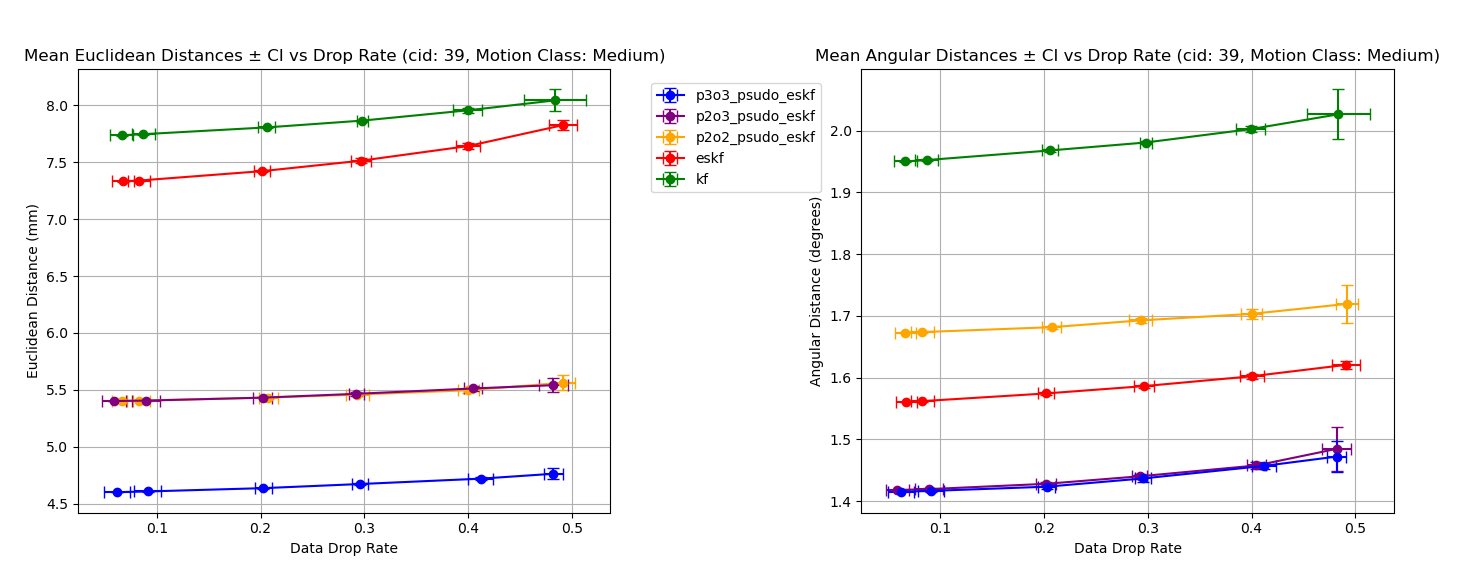}
        \includegraphics[width=\columnwidth]{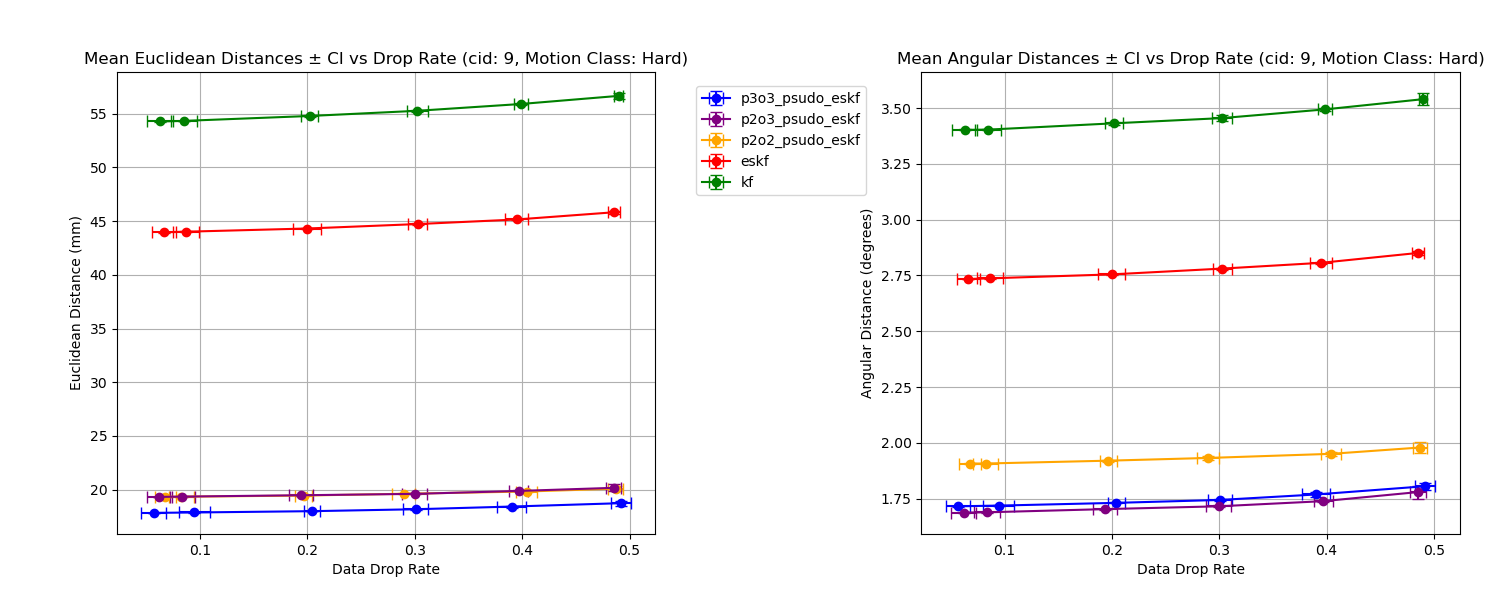}
        \caption{Prediction error (mean) with different packet loss rates for (top)
        easy, (middle) medium, and (bottom) hard motion classes.}
        \label{fig:prediction_error_packet_loss}
    \end{figure}

    To ensure rigorous validation, packet loss is simulated by generating random
    floating-point numbers uniformly distributed in the interval (0, 1); a
    packet is considered received if the generated number exceeds the specified
    drop rate, and dropped otherwise. For each predictor and each drop rate, the
    experiment is repeated at least 10 times. The confidence intervals for both
    position (Euclidean distance) and orientation (angular distance) errors are
    then computed and visualized as error bars at each data point in the diagram.

    Figure~ \ref{fig:prediction_error_packet_loss} demonstrates the robustness of
    the proposed PsudoESKF method compared to the baseline KF and ESKF under varying
    packet loss rates across different motion classes. As packet loss increases,
    all methods experience degradation in prediction accuracy; however, \texttt{p3o3\_PsudoESKF}
    not only achieves the lowest overall error in both position and orientation,
    but also exhibits a much smaller increase in error (i.e., a gentler slope) as
    packet loss rises compared to the other methods. This highlights both the
    superior accuracy and the enhanced robustness of p3o3\_PsudoESKF under
    challenging network conditions, which are critical for real-time VR applications
    where network instability is common.

    As for different patterns in Figure~ \ref{fig:prediction_error_packet_loss},
    the mean orientation error for the p3o3 variant is reduced by approximately 37\%
    compared to KF for easy motion patterns and by about 49.6\% for hard motion
    patterns at a 50\% packet loss rate. The mean position error for the p3o3 variant
    is reduced by approximately 50.8\% compared to KF for easy motion patterns and
    by about 66.1\% for hard motion patterns at a 50\% packet loss rate. This highlights
    the effectiveness of the PsudoESKF method in maintaining prediction accuracy
    even under challenging network conditions. As the motion pattern becomes
    more unpredictable, methods that incorporate the highest-order derivative integration
    (p3o3) demonstrates even better performance for both positional and orientational
    prediction against packet loss.



    \section{Discussion}
    The results show that the proposed method outperforms the baseline methods in
    terms of prediction accuracy and robustness to data loss. The proposed
    method achieves lower prediction errors for both position and orientation
    across different motion patterns, indicating its effectiveness in handling various
    user movements. The results also demonstrate that the proposed method
    maintains the lowest prediction error across prediction horizons up to 100
    ms.

    Although the proposed method achieves lower prediction errors than the
    baselines, it is important to recognize that metrics such as Mean Squared Error
    (MSE), Absolute Trajectory Error (ATE) and Relative Pose Error (RPE) are not
    sufficient to fully characterize predictor performance in XR applications.
    MSE quantifies average squared differences between predicted and ground-truth
    values, while ATE and RPE assess overall trajectory alignment and local consistency,
    respectively. However, these metrics primarily capture average accuracy and
    do not account for perceptual artifacts such as jitter or short-term
    instability, which can significantly affect user experience~\cite{Jiang_Qinjun_2025}.
    Therefore, future evaluations will incorporate user-centric metrics that better
    reflect perceptual quality and comfort, ensuring that improvements in
    trajectory accuracy is being translated to tangible benefits in the XR user
    experience.


    Building upon the preceding analysis of limitations and challenges, we now
    consider potential industrial applications and broader implications of the
    proposed predictor. Beyond remote XR, the proposed predictor can be
    beneficial in teleoperation scenarios, such as FPV for drones. This
    framework is particularly valuable in applications requiring precise
    navigation and control, such as search and rescue operations, industrial
    inspections, or recreational drone activities. By leveraging accurate
    trajectory prediction, the proposed method can enhance the alignment between
    user input and drone motion, ensuring smoother navigation and reducing the
    risk of collisions. This capability is especially critical in environments with
    limited visibility or high-speed operations, where precise and responsive
    control is essential for achieving mission objectives effectively. Accurate trajectory
    prediction helps ensure smooth navigation and obstacle avoidance. By
    improving trajectory alignment, the framework can enhance user control,
    reducing collision risks and supporting applications like search and rescue,
    industrial inspections, and recreational drone activities, where precision and
    responsiveness are critical.

    \section{Conclusion}
    In this paper, we propose a context-aware motion prediction framework for
    head-mounted displays in latency-sensitive virtual reality. Our main contribution
    is the PsudoESKF, a lightweight ESKF that incorporates higher-order motion
    modelling and an entropy-based motion classifier. We showed that PsudoESKF consistently
    outperforms standard KF and ESKF baselines in both accuracy and robustness,
    particularly for unpredictable motion and under network packet loss. Importantly,
    our method requires only pose data and is efficient for edge deployment. These
    results demonstrate that combining higher-order dynamics with context-awareness
    provides a practical and effective solution for reducing MTP latency to enable
    better user experiences in offloaded XR applications.

    \bibliographystyle{ACM-Reference-Format}
    \bibliography{reference}









\end{document}